\documentstyle[preprint,tighten,version2,aps]{revtex}

{\makeatletter%
\gdef\labeleqs#1{{%
\edef\@currentlabel{%
\ifappendixon\appletter\fi
\ifsecnumbers\ifnum\c@secnum>0
\arabic{secnum}.\fi\fi\arabic{equation}}%
\label{#1}%
}}%
}
\begin{document}
\draft
\preprint{IFUP-TH 64/94}
\begin{title}
Strong coupling analysis of the large-$N$ 2-d lattice chiral models.
\end{title}
\author{Massimo Campostrini, Paolo Rossi, and Ettore Vicari}
\begin{instit}
Dipartimento di Fisica dell'Universit\`a and I.N.F.N.,
I-56126 Pisa, Italy
\end{instit}
\begin{abstract}
Two dimensional $N=\infty$ chiral models on the square and honeycomb lattices
are investigated by a strong coupling analysis.
Strong coupling expansion turns out to be predictive for  the evaluation
of continuum physical quantities, to the point of showing asymptotic scaling.
Indeed in the strong coupling region a quite large range of $\beta$ values
exists
where the fundamental mass agrees, within about 5\% on the square lattice
and about 10\% on the honeycomb lattice, with the continuum predictions in the
energy scheme.

\end{abstract}


\narrowtext

\section{Introduction}

Recent numerical studies of lattice
two-dimensional ${\rm SU}(N)\times{\rm SU}(N)$ principal chiral models,
with the standard nearest-neighbour interaction
\begin{equation}
S_L = -2 N \beta \sum_{x,\mu}
{\mathop{\rm Re}}\,{\mathop{\rm Tr}} \,[ U(x)\,U^\dagger(x{+}\mu) ]\;,
\;\;\;\;\;\;\;\;\;\;\;\;\;\beta\;=\;{1\over NT}\;,
\end{equation}
have shown the existence of a scaling region, where continuum
predictions for dimensionless ratios of physical quantities are
substantially verified \cite{chiral1,chiral2}.
The scaling region begins at relatively small values of the correlation
length well within the expected region of convergence of
strong-coupling expansion.  Moreover by performing a
variable change~\cite{Parisi} from the temperature $T$ to
\begin{equation}
T_E = {8N \over N^2 - 1 }E\;,
\;\;\;\;\;\;\;\;\;\;\;\;\;\beta_E\;=\;{1\over NT_E}\;,
\label{betaE}
\end{equation}
where $E$ is the internal energy,
one can find agreement
in the whole
scaling region between the measured mass scale and the
asymptotic scaling prediction, within few per cent~\cite{chiral2}.

As a matter of fact, this may be considered  as an
evidence for asymptotic scaling within the strong-coupling regime,
motivating a test of scaling and asymptotic scaling by  strong coupling
computations.
As a byproduct, strong-coupling series can be analyzed to investigate
the critical behavior of the $N=\infty$ theory, where Monte Carlo data
seem to indicate the existence of a phase transition at finite $\beta$.

Ref.~\cite{SC} was devoted to a complete presentation of our strong coupling
calculations
performed by means of the character expansion.
We calculated strong coupling series for several quantities
on the square and honeycomb lattices.
On the ordinary square lattice,
we calculated the free energy up to $O\left( \beta^{18}\right)$,
and the fundamental Green's function
\begin{equation}
G(x)\;=\;\langle \;{1\over N} {\rm Re} \;{\rm Tr} \,[ U(x) U(0)^\dagger
]\;\rangle\;\;\;
\end{equation}
up to $O\left( \beta^{15}\right)$.
For chiral models on the honeycomb lattice,
defined by the nearest-neighbour action,  longer series were obtained:
the free energy up to $O\left( \beta^{26}\right)$ and $G(x)$ up to
$O\left( \beta^{20}\right)$.
Lattice chiral models on square and honeycomb lattices are expected
to belong to the same class of universality with respect to the continuum
limit.
As we will see from the strong coupling analysis,
even at finite $\beta$ large-$N$ chiral models on the honeycomb lattices
show a pattern very similar to that observed on the square lattice.

In this paper, which represents the logical continuation of Ref.~\cite{SC},
we analyze the $N=\infty$ strong coupling series presented there
and the results are compared with
the continuum limit predictions and Monte Carlo simulations.
The main result of our strong coupling analysis of 2-d $N=\infty$ chiral models
on the square and honeycomb lattices is the identification of a scaling region
where
known continuum results are reproduced with good accuracy, and asymptotic
predictions are
substantially fulfilled in the energy scheme.

\section{Strong coupling evidence of a large-$N$ phase transition.}

Numerical simulations at large $N$ of $SU(N)$ and $U(N)$ lattice chiral models
show evidence of a phase transition at $N=\infty$. Indeed sharper
and sharper peaks in the specific heat
\begin{equation}
C\;=\;{1\over N} {{\rm d} E\over {\rm d}T}
\end{equation}
are observed with increasing $N$,
suggesting a divergent large-$N$ limit at a finite $\beta$~\cite{chiral2}.
By extrapolating to $N=\infty$ the positions of the specific heat peaks,
we obtained  a rather precise estimate of the critical coupling:
$\beta_c=0.3057(3)$.
Details on our Monte Carlo simulations and their analysis can be found in
Ref.~\cite{LNPH}.

In order to investigate the above issue, we analyze the $N=\infty$
strong coupling series by employing the integral approximant
technique~\cite{Hunter,Fisher},
which is especially recommended in the case of small critical
exponent~\cite{Guttmann}.
The method of integral
approximants consists of representing the power series under study
by the integral of a linear differential equation.
In our analysis we considered the integral approximants obtained from a first
order linear differential equation
\begin{equation}
Q_L(x)f^\prime (x)\,+\, P_K(x)f(x)\,+\,R_J(x)\;=\; O\left(
x^{L+K+J+2}\right)\;,
\label{intapprx}
\end{equation}
where $Q_L$,$P_K$ and $R_J$ are respectively  $L,K$, and $J$ order
polynomials (we fix $Q_{L,0}=1$).
These approximants are singular at the zeroes $x_0$ of $Q_L(x)$, and behave as
\begin{equation}
A(x)\,|x-x_0|^{-\gamma}\,+\,B(x)\;,
\label{intapprx2}
\end{equation}
where $A(x)$ and $B(x)$ are regular in the neighbourhood of $x_0$, and
\begin{equation}
\gamma \;=\; - {P_K(x_0)\over Q^\prime_L(x_0)}\;.
\end{equation}
Given a $M$ order series, $L,K$ and $J$ must satisfy the condition
$L+K+J+2 \leq M\;$.

Let us analyze the $N=\infty$ strong coupling series of the specific heat,
which is even in $\beta$:
\begin{eqnarray}
\beta^{-2}C\;=&&1+6\,x+30\,x^2+266\,x^3+2160\,x^4+
19932\,x^{5}\nonumber \\
&& + 183638\,x^{6} + 1754130\,x^{7}+16911192\,x^{8}+
O\left(x^{9}\right)\;,
\label{Cseries}
\end{eqnarray}
where $x=\beta^2$.
In Table~\ref{Cresum} we report the first singularity
in the real axis and the corresponding
exponent for different values of $L,K,$ and $J$.
The results are quite stable, leading to a critical behavior of the specific
heat
typical of a second order phase transition:
\begin{equation}
C\;\sim\; |\beta - \beta_c|^{-\alpha}\;.
\label{critC}
\end{equation}
{}From Table~\ref{Cresum} we estimate
\begin{eqnarray}
\beta_c&=&0.3058(3)\;,\nonumber\\
\alpha&=& 0.23(3) \;.
\end{eqnarray}
The errors are just indicative. They are the variance
of the results in Table~\ref{Cresum} after discarding the two furthest values
from the corresponding average; they should give an idea of the
spread of the results coming from different approximants.
Notice that the strong coupling determination  of $\beta_c$ is in agreement
with its estimate
from numerical simulations at large $N$.

As further check of the above resummation procedure,
Fig.~\ref{C_largeN} compares, in the region $\beta < \beta_c$,
$SU(N)$ and $U(N)$ Monte Carlo data of the specific heat at large $N$
($N=21,30$ for $SU(N)$, and $N=15,21$ for $U(N)$)
with the determinations
coming from the resummed and the plain strong coupling series (\ref{Cseries}).
We recall that $SU(N)$ and $U(N)$ models should have the same large-$N$ limit.
Monte Carlo data of $C$ appear to approach, for growing $N$, the
determination from the resummed strong coupling series.
As expected from simple considerations on the finite $N$ corrections to the
$N=\infty$
strong coupling series, $U(N)$ models converge
faster than $SU(N)$ models to the $N=\infty$ limit in the strong coupling
region.

Monte Carlo data at large $N$ seem to indicate that
all physical quantities, such as the magnetic susceptibility
$\chi\equiv \sum_x G(x)$ and the second moment mass $M^2_G$ ($M_G\equiv
1/\xi_G$ and
$\chi\xi^2_G\equiv {1\over 4}\sum_x x^2 G(x)\;$),
are well behaved functions of the internal energy even at
$N=\infty$~\cite{chiral2}.
Therefore, as a consequence of the specific heat divergence,
$\chi$ and $M^2_G$ should have a  singular behavior with respect to $\beta$.
We indeed expect
\begin{equation}
{{\rm d}\ln \chi\over {\rm d}\beta}\;\sim \;
{{\rm d}\ln M^2_G\over {\rm d}\beta}\;\sim\;|\beta-\beta_c|^{-\alpha}\;.
\label{dchicrit}
\end{equation}
in the neighbourhood of $\beta_c$.
Notice that a behavior like Eq.~(\ref{dchicrit}) leads to a non-analytical zero
of the
$\beta$-function $\beta_L(T)$ at $\beta_c$:
\begin{equation}
\beta_L(T) \;\sim\; |\beta-\beta_c|^\alpha
\label{betafcrit}
\end{equation}
around $\beta_c$, explaining the observed behavior with respect to  $\beta$
of the large-$N$ Monte Carlo data for the fundamental mass~\cite{chiral2}.

In order to check the behavior (\ref{dchicrit}), we analyzed the corresponding
strong coupling series by a modified integral approximant
scheme forcing the approximant to have a singularity at $\beta\simeq 0.3058$,
obtaining biased estimates of the exponent in Eq.~(\ref{dchicrit}).
In this modified scheme  the values of $L,K$, and $J$ in Eq.~(\ref{intapprx})
must be
chosen according to  the condition
$L+K+J+1\leq M$.
We analyze the series
\begin{eqnarray}
{{\rm d}\ln \chi\over {\rm
d}\beta}\;=&&4+8\beta+28\beta^2+48\beta^3+204\beta^4+440\beta^5
+1740\beta^6+3744\beta^7+15148\beta^8 +35048\beta^9\nonumber \\
&&+140980\beta^{10}+327600\beta^{11}
+1323612\beta^{12}+3149112\beta^{13}+12727908\beta^{14}+
O\left( \beta^{15}\right)\;,
\label{dlogchi}
\end{eqnarray}
and
\begin{eqnarray}
{{\rm d}\ln \beta M^2_G\over {\rm d}\beta}\;
=&&-4-10\beta-28\beta^2-74\beta^3-224\beta^4-598\beta^5
-1936\beta^6-5282\beta^7-17560\beta^8\nonumber \\
&&-49170\beta^9-162144\beta^{10}-464426\beta^{11}
-1549656\beta^{12}-4459234\beta^{13}+
O\left( \beta^{14}\right)\;.
\label{dlogMg}
\end{eqnarray}
In Table~\ref{chiresum} we report the range of the exponent variations
when varying the zero in the interval $0.3055-0.3061$.
These results are quite consistent with the exponent $\alpha$ obtained in the
analysis of the specific heat strong coupling series,
supporting the relations (\ref{dchicrit}) and therefore (\ref{betafcrit}).
When performing an unbiased analysis of the series (\ref{dlogchi}) and
(\ref{dlogMg}), that is
without forcing the approximants to have a zero
at a fixed $\beta$, the singularity and the corresponding exponent
turn out to be less stable; more terms in the series would be necessary to have
a satisfactory analysis independent of that of the specific heat.

In Fig.~\ref{magsusc} we compare our strong coupling calculations
of $\chi$ with the Monte Carlo data of $SU(N)$ and $U(N)$ models at large $N$.
An improved strong  coupling estimate of $\chi$, represented by the full line
in Fig.~\ref{magsusc},
was obtained  by integrating the resummed
series of ${\rm d}\ln \chi/{\rm d}\beta$
(using $L=5$, $K=4$ and $J=4$, see Table~\ref{chiresum}).
Growing $N$, the $SU(N)$ and $U(N)$ data of $\chi$ and $\xi_G$ approach the
same $N=\infty$ limit, which is well reproduced  by the resummation of the
series
(\ref{dlogchi}) and (\ref{dlogMg}) in the strong coupling region.

We should mention an apparent discrepancy between
the $SU(N)$ Monte Carlo results at large $N$ and the resummation of the
$N=\infty$ series.
In the $SU(N)$ numerical simulations, when varying $N$ (for sufficiently large
$N$)
the position of the peak of the specific heat turns out to be quite stable
with respect to the correlation length, $\xi_G^{peak}\simeq 2.8$, leading to
the expectation
that at the large-$N$ critical point $\xi_G^{(c)}\simeq 2.8$~\cite{chiral2}.
On the other hand the estimate of $\xi_G$ from the resummation of the strong
coupling series
is slightly smaller: $\simeq 2.5$.

\section{Scaling and asymptotic scaling.}

In spite of the existence of a phase transition at $N=\infty$, Monte
Carlo data at large $N$ showed scaling and approximate
asymptotic scaling (in the energy
scheme) even for $\beta$ smaller then the peak of the specific
heat~\cite{chiral2}.
The stability of this pattern suggests
an effective decoupling of the modes responsible for the
phase transition from those determining the physical continuum limit,
and therefore that evidences of scaling and asymptotic scaling could be
provided
by the large-$N$ strong coupling expansion.

The on-shell fundamental mass $M$ can be  extracted from the long distance
behavior
of the correlation function in the fundamental channel
$G(x)$, or from the imaginary  pole of its Fourier transform.
We considered two estimators of $M$, $\mu_{\rm s}$ and $\mu_{\rm d}$, defined
from the long distance
behavior  of wall-wall correlation functions constructed
with $G(x)$ respectively along the sides and the diagonals of the lattice.
An alternative mass $M_G$ is defined from the inverse second moment of $G(x)$.
Unlike the on-shell mass $M$, $M_G$ is an off-shell quantity, it is related
to the zero momentum of the Fourier transform of $G(x)$, indeed
${\widetilde G}(p)^{-1}\sim M^2_G+p^2$ at small momentum.

The quantities $\mu_{\rm s}$, $\mu_{\rm d}$ and $M^2_G$
enable us  to perform tests of scaling based on rotational invariance
at distances $d \gtrsim \xi\equiv 1/\mu_{\rm s}$,
checking $\mu_{\rm s}/\mu_{\rm d}\simeq 1$, and on the stability of
dimensionless physical quantities,
looking at the ratio $\mu_{\rm s}/M_G$.
We should say that these tests concern the long distance physics of chiral
models.

Monte Carlo data at relatively large $N$ showed that, within statistical errors
of few per mille, the above scaling requirements are verified already at
$\xi\simeq \xi_G\simeq 2$,
well within the strong coupling region.
Numerical simulations provided an estimate of the large-$N$ limit
of the ratio $M/M_G$ in the continuum limit: $M/M_G=0.991(1)$~\cite{chiral2}.

In Ref.~\cite{SC} the strong coupling series corresponding to the above
mentioned quantities have been calculated, in particular
$M^2_G$ up to $O\left( \beta^{13}\right)$,
$M_{\rm s}^2\equiv 2(\cosh\mu_{\rm s} - 1)$ up to $O\left( \beta^{11}\right)$,
and $M_{\rm d}^2\equiv 4(\cosh\mu_{\rm d}/\sqrt{2} - 1)$ up to $O\left(
\beta^{10}\right)$.

In Fig.~\ref{rotinv} we plot the ratio $\mu_{\rm s}/\mu_{\rm d}$ vs. the
correlation length $\xi_G\equiv 1/M_G$ as obtained from our strong coupling
series.
The $N=\infty$ strong coupling curve confirms the large-$N$ Monte Carlo result:
$\mu_{\rm s}/\mu_{\rm d}\simeq 1$ within few  per mille at $\xi\simeq 2$.

Fig.~\ref{scaling} shows the ratio $\mu_{\rm s}/M_G$ vs. $\xi_G$.
Notice the stability of the curve for a large region of values of $\xi_G$
and the good agreement (well within 1\%) with the continuum large-$N$
value extrapolated by Monte Carlo data.

In order to test asymptotic scaling we perform the variable change
indicated in Eq.(\ref{betaE}), evaluating the energy from its
strong-coupling series.
The two loop renormalization group and a Bethe Ansatz evaluation
of the mass $\Lambda$-parameter ratio~\cite{Balog}
lead to the following large-$N$ asymptotic scaling prediction for
the on-shell fundamental mass in the $\beta_E$ scheme:
\begin{eqnarray}
&&M \;\cong\;
R_E\,
\Lambda_{E,2l}(\beta_E)\;,\nonumber \\
&&R_E\;=\;
16\sqrt{{\pi\over e}} \exp \left({\pi\over 4}\right)\;,\nonumber \\
&&\Lambda_{E,2l}(\beta_E) \;=\;
\sqrt{8\pi\beta_E}\exp(-8\pi\beta_E)\;,\nonumber \\
&&\beta_E\;=\;{1\over 8E}\;.
\label{mass-lambda}
\end{eqnarray}
In Fig.~\ref{asyscaling} the strong-coupling estimates of
$\mu_{\rm s}/\Lambda_{E,2l}$ and $M_G/\Lambda_{E,2l}$ are plotted vs.\
$\beta_E$, for a region of coupling corresponding to correlation
lengths $1.5\lesssim \xi_G \lesssim 3$.
(We recall that $M_G$ differs from $M$ by about 1\% in the continuum limit.)
The agreement with the
exact continuum prediction is within about 5\% in the whole region.
Notice also that both curves go smoothly through the value of $\beta_E$
corresponding
to the specific heat singularity $\beta_c$, which is $\beta_E^{(c)}\simeq
0.220$.

The strong coupling curves in Fig.~\ref{asyscaling} were obtained from
the plain series of the energy and respectively of $M_{\rm s}^2$ and $M^2_G$.
In the case of $M_G$, we also determined $M_G/\Lambda_{E,2l}$
evaluating the energy and $M^2_G$
by integrating the resummed series respectively of the specific heat
and of ${\rm d}\ln \beta M^2_G/{\rm d}\beta$. The resulting curve changes
very little from that derived from the plain series, the difference
between the two curves would not be visible in Fig.~\ref{asyscaling}.
This indicates once more that the change of variable $\beta\rightarrow \beta_E$
washes out the singularity in $\beta$ when considering physical quantities.

\section{Chiral models on the honeycomb lattice.}

On the honeycomb lattice we consider the action with
nearest-neighbour interaction.
It can be written as a sum over all links of the honeycomb lattice:
\begin{equation}
S_{\rm h} = -2 N \beta \sum_{\rm links}
{\mathop{\rm Re}}\,{\mathop{\rm Tr}}\,[ U_l\,U^\dagger_r]\;,
\;\;\;\;\;\;U\in SU(N)\;,
\label{exaction}
\end{equation}
where $l,r$ indicate the sites at the ends of each link.
As on the square lattice, a lattice space $a$,
which represents the lattice length unit, is defined to be the length
of a link. The volume of an hexagon is $v_{\rm h}=3\sqrt{3}/2$.
Straightforward calculations show that the correct continuum limit is obtained
identifing
\begin{equation}
T\;=\; {\sqrt{3}\over N\beta}\;.
\label{tempex}
\end{equation}

\subsection{The large-$N$ phase transition.}

On the honeycomb lattice
we have calculated the $N=\infty$ strong coupling series of the specific heat,
which is even in $\beta$, up to $26^{\rm th}$ order in $\beta$:
\begin{eqnarray}
\beta^{-2}C\;=&&1+10\,x^2+90\,x^4+
396\,x^{5}+728\,x^{6}+9120\,x^{7}+28186\,x^{8}+ 136800\,x^{9}\nonumber \\
&&  + 886116\,x^{10}+3129380\,x^{11}+
+18935800\,x^{12}+O\left(x^{13}\right)\;,
\label{Cseriesex}
\end{eqnarray}
where $x=\beta^2$.
The integral approximant analysis of the above series, whose results are
reported in
Table~\ref{Cresumex}, leads again to a second order type critical behavior with
the following estimates of the critical $\beta$ and $\alpha$ exponent:
\begin{eqnarray}
\beta_c&=&0.4339(1)\;,\nonumber\\
\alpha&=& 0.17(1) \;.
\end{eqnarray}
Notice that this estimate of the exponent $\alpha$ is very close to that of the
square lattice. The uncertainty on both estimates cannot really exclude
that they are equal, which would be an indication of universality.

Also in this context we analyzed the strong coupling series of
the logarithmic derivative of the magnetic susceptibility
$\chi$ and $\beta M^2_G$ by the modified integral
approximant method which forces the existence of a zero at $\beta_c$.
In Table~\ref{chiresum} we report the range of the exponent variations
when varying the zero in the interval $0.4338-0.4340$.
As for the square lattice,
the results in Table~\ref{chiresum} are consistent with a divergence
characterized by the specific heat exponent (cfr. Eq.~(\ref{dchicrit}),
supporting the existence of a non-analytical zero
of the $\beta$-function at $\beta_c$.

\subsection{Scaling.}

On the hexagonal lattice
the maximal violation of the full rotational symmetry occurs for
directions differing by a $\pi/6$ angle, and therefore,
taking into account its discrete rotational symmetry, also by
a $\pi/2$ angle. So a good test of rotation invariance is provided by the
ratio between masses extracted from the long distance
behaviors of a couple of orthogonal wall-wall correlation functions constructed
with $G(x)$.

In Ref.~\cite{SC} we defined two orthogonal wall-wall correlation functions
$G^{{\rm(w)}}_1(x)$  and $G^{{\rm(w)}}_2(x)$, with the corresponding masses
$\mu_1$ and $\mu_2$, which should both reproduce
the on-shell fundamental mass $M$ in the continuum limit.
In order to extract $\mu_1$ and
$\mu_2$ we evaluated the $O\left( \beta^{18}\right)$
series of $\exp (-3\mu_1/2)$
and the $O\left( \beta^{17}\right)$ series of
$\exp (-\sqrt{3}\mu_2/2)$.

Fig.~\ref{rotinvex} shows the ratio $\mu_1/\mu_2$ vs. $\xi_G\equiv 1/M_G$.
As expected from the better rotational symmetry of the honeycomb lattice,
rotation invariance is set earlier than for the square lattice:
already at a correlation length $\xi_G\simeq 0.5$ $\mu_1/\mu_2\simeq 1$
within 1\%.

In Fig.~\ref{scalingex} we plot the ratio $\mu_1/M_G$ vs. $\xi_G$.
The approach to the continuum limit value seems to be substantially
equivalent to that observed on the square lattice,
but then for $\xi_G \gtrsim 1.5$ the curve becomes unstable.
Such an instability should be cured by an extension of the series.

\subsection{Asymptotic scaling.}

The asymptotic scaling test is again best performed in the energy scheme.
This requires some weak coupling calculations, which present some subtleties
on the honeycomb lattice.
This is essentially due to the fact that, unlike square and triangular
lattices,  lattice sites are not characterized by a group of translations.
Details on our weak coupling calculations are given in the Appendix.

We calculated the internal energy (per link) up to two loops, finding
\begin{equation}
E\;=\; {N^2-1\over N}\,
{T\over 6\sqrt{3}}\;\left[ 1\;+\;{N^2-2\over N}\,
{T\over 24\sqrt{3}}\;+\;O\left( T^2\right)\right]\;.
\label{energyex}
\end{equation}
The energy scheme consists in defining a new temperature $T_E$
proportional to the energy
\begin{equation}
T_E\;=\; {6\sqrt{3}N\over N^2-1}E \;,\;\;\;\;\;\;\;
\beta_E\;=\; {1\over NT_E}\;.
\label{tex}
\end{equation}

The other important ingredient in this game is the mass $\Lambda$-parameter
ratio in the honeycomb lattice regularization, which requires
the calculation of the ratio between the $\Lambda$-parameter of
the ${\overline {MS}}$ renormalization scheme $\Lambda_{\overline {MS}}$
and that of the honeycomb lattice regularization $\Lambda_{\rm h}$,
given that the the (on-shell) mass $\Lambda$-parameter ratio in the
${\overline {MS}}$ scheme is known~\cite{Balog}.

{}From a one loop calculation we obtained
\begin{equation}
{\Lambda_{\overline {MS}}\over \Lambda_{\rm h}}\;=\;
4\exp \left({N^2-2\over N^2}\,{2\pi\over 3\sqrt{3}} \right)\,\;.
\label{ratio1}
\end{equation}

The ratio between $\Lambda_{{\rm h},E}$, the $\Lambda$ parameter of the energy
scheme, and $\Lambda_{\rm h}$ is easily obtained from the two loop term
of the internal energy:
\begin{equation}
{\Lambda_{\rm h}\over \Lambda_{{\rm h},E}}\;=\;
\exp \left(- {N^2-2\over N^2}\,{\pi\over 3\sqrt{3}}\right)\, \;.
\label{ratioe}
\end{equation}

Then the $N=\infty$ asymptotic scaling prediction
in the energy scheme is
\begin{eqnarray}
&&M \;\cong\;
R_{{\rm h},E}\,
\Lambda_{E,2l}(\beta_E)\;,\nonumber \\
&&R_{{\rm h},E}\;=\;
8\sqrt{{2\pi\over e}} \exp \left( {\pi\over 3\sqrt{3}}\right)\;,\nonumber \\
&&\Lambda_{E,2l}(\beta_E) \;=\;
\sqrt{8\pi\beta_E}\exp(-8\pi\beta_E)\;,\nonumber \\
&&\beta_E\;=\;{1\over 6\sqrt{3}E}\;.
\label{mass-lambdaex}
\end{eqnarray}

Fig.~\ref{asyscalingex} shows the ratios
$\mu_1/\Lambda_{E,2l}$ and $M_G/\Lambda_{E,2l}$ vs.\
$\beta_E$ (corresponding to correlation lengths $1\lesssim \xi_G\lesssim 2.5$),
as obtained from the corresponding strong coupling series.
Again there is good agreement with the continuum prediction,
especially in the region corresponding to correlation length
$\xi_G\gtrsim 2$, where the agreement is within 10\%.
The curve corresponding to $M_G$ is more stable, and
it changes little when calculated resumming the involved series.

\appendix{Weak coupling expansion on the honeycomb lattice.}

On the honeycomb lattice the sites cannot be associated to a group of
translation.
This causes a few subtleties in the analysis of models on such a lattice.

The sites $\vec{x}$ of a finite periodic hexagonal lattice can be represented
in cartesian coordinates by
\begin{eqnarray}
&&\vec{x}(l_1,l_2,l_3) \;=\;
l_1\,\vec{\eta}_1\,+\,l_2\,\vec{\eta}_2\,+\,l_3\left({1\over 2},0\right)\;,
\nonumber \\
&&l_1=1,...L_1\;,\;\;\;\;l_2=1,...L_2\;,\;\;\;\;l_3=-1,1\;,\nonumber \\
&&\vec{\eta}_1 \;=\; \left({3\over 2},{\sqrt{3}\over 2}\right)\;,
\;\;\;\;\;\;\;\;\;\;
\vec{\eta}_2 \;=\; \left(0,\sqrt{3}\right)\;.
\label{a1}
\end{eqnarray}
We set $a=1$,  where the lattice space $a$ is the length of a link.
The total number of exagons on the lattice is $L_1L_2$, while the sites
are $2L_1L_2$.
The coordinate $l_3$ can be interpreted as the parity of the corresponding
lattice site: sites with the same parity are connected by an even number of
links.

Notice that each of the two sublattices identified by
$\vec{x}_-(l_1,l_2)\equiv\vec{x}(l_1,l_2,-1)$ and $\vec{x}_+(l_1,l_2)\equiv
\vec{x}(l_1,l_2,1)$ forms a triagular lattice.
Each link of the honeycomb lattice connects sites belonging to different
sublattices.
Triangular lattices have a more symmetric structure, in that their sites are
characterized by a group of translations.
It is then convenient to rewrite a field $\phi(\vec{x})\equiv\phi(l_1,l_2,l_3)$
in terms of two new fields $\phi_-(\vec{x}_-)\equiv \phi(\vec{x}_-)$
and $\phi_+(\vec{x}_+)\equiv \phi(\vec{x}_+)$ defined respectively on the
sublattices
$\vec{x}_-$ and $\vec{x}_+$.
A finite lattice Fourier transform can be consistently defined
\begin{eqnarray}
&&\phi_\pm(\vec{p})\;=\;v_{\rm h}\,\sum_{\vec{x}_\pm} e^{i\vec{p}\cdot
\vec{x}_\pm}\,\phi_\pm (\vec{x}_\pm)\;,\nonumber \\
&&\phi_\pm(\vec{x}_\pm)\;=\;{1\over v_{\rm h}L_1L_2}\,\sum_{\vec{p}}
e^{-i\vec{p}\cdot
\vec{x}_\pm}\,\phi_\pm (\vec{p})\;,
\label{a3}
\end{eqnarray}
where $v_{\rm h}=3\sqrt{3}/2$ is the volume of an hexagon, and
the set of momenta is
\begin{eqnarray}
&&\vec{p}\;=\;{2\pi\over L_1}m_1\vec{\rho}_1\;+\;{2\pi\over L_2}m_2\vec{\rho}_2
\nonumber \\
&&m_1=1,...L_1\;,\;\;\;\;m_2=1,...L_2\;,\nonumber \\
&&\vec{\rho}_1\;=\;\left( {2\over 3},0\right)\;,\;\;\;\;\;\;\;
\vec{\rho}_2\;=\;\left( -{1\over 3},{1\over \sqrt{3}}\right)\;.
\label{a4}
\end{eqnarray}
Notice that
\begin{equation}
\vec{p}\cdot\vec{x}\;=\; {2\pi\over L_1}l_1m_1\,+\,{2\pi\over L_2}l_2m_2
\,+\,l_3{p_1\over 2}\;.
\label{a5}
\end{equation}

To begin with, let us discuss the simple Gaussian models, whose action can be
written
as
\begin{equation}
S_{\rm G}\;=\;{\kappa\over 2}\sum_{\rm links}\left(\phi(x_l)-\phi(x_r)\right)^2
\label{a6}
\end{equation}
where $x_l,x_r$ indicate the sites at the ends of each link.
Rewriting the field $\phi(x)$ in terms of two fields $\phi_-(x_-)$
and $\phi_+(x_+)$ as described above, and performing the Fourier transform
(\ref{a3}) we obtain
\begin{eqnarray}
S_{\rm G}\;=&&{\kappa \over \sqrt{3}}{1\over v_{\rm h}L_1L_2}\sum_p [
\phi_-(-p)\phi_-(p) + \phi_+(-p)\phi_+(p)\nonumber \\
&&\;\;-\phi_-(-p)\phi_+(p)H(-p) - \phi_+(-p)\phi_-(p)H(p)]\;,
\label{a7}
\end{eqnarray}
where
\begin{equation}
H(p)\;=\;  e^{-ip_1}{1\over 3}\left(
1\,+\, 2e^{i{3p_1\over 2}}\cos {\sqrt{3}p_2\over 2}\right)\;.
\label{a8}
\end{equation}
{}From (\ref{a7}) we derive the propagators:
\begin{eqnarray}
&&\left< \phi_-(k)\phi_-(q)\right>\;=\;\left< \phi_+(k)\phi_+(q)\right>\;=\;
v_{\rm h}{\sqrt{3}\over \kappa} {1\over \Delta(k)}\,\delta_{k+q,0} \;,
\nonumber \\
&&\left< \phi_+(k)\phi_-(q)\right>\;=\;
v_{\rm h}{\sqrt{3}\over \kappa} {H(k)\over \Delta(k)}\,\delta_{k+q,0} \;,
\label{a9}
\end{eqnarray}
where
\begin{equation}
\Delta(k)\;=\; {8\over 9} \left[
2\,-\,\cos {\sqrt{3}\over 2}k_2\left( \cos {3\over 2}k_1+
\cos {\sqrt{3}\over 2}k_2\right)\right]\;.
\label{a10}
\end{equation}
When $x_+$ and $x_-$ are the ends of the same link, i.e.
$|x_+-x_-|=1$, one can easily prove that
\begin{equation}
\left< \phi_+(x_+)\phi_-(x_-)\right>
 - \left< \phi_-(x_-)\phi_-(x_-)\right>
\;=\; -{1\over 3\kappa}\;.
\label{a11}
\end{equation}

The nearest-neighbour action of chiral models on the honeycomb lattice is
\begin{equation}
S_{\rm h}\;=\;-{\sqrt{3}\over T}\sum_{\rm links}
2{\mathop{\rm Re}}\,{\mathop{\rm Tr}}\,[ U_l\,U^\dagger_r]\;,
\;\;\;\;\;\;U\in SU(N)\;,
\label{a12}
\end{equation}
where $l,r$ indicate the sites at the ends of each link.
The perturbative expansion is performed by setting
\begin{equation}
U\;=\;e^{iA}\;\;, \;\;\;\;\;A=\sum_a T_a A_a\;\;,
\label{a13}
\end{equation}
($T_a$ are the generators of the $SU(N)$ group and $A_a$ are
$N^2-1$ real fields) and expanding $U$ in powers of $A$.
The action $S_{\rm h}$ becomes
\begin{equation}
S_{\rm h}\;=\;{\sqrt{3}\over T} \sum_{\rm links}
\left[ {\rm Tr} (A_l-A_r)^2 \;+\;
{1\over 4}{\rm Tr} (A_l^2-A^2_r)^2 \;-\;
{1\over 3}{\rm Tr} (A_l-A_r)(A_l^3-A_r^3)\;+\; O\left( A^6\right)\right]\;.
\label{a14}
\end{equation}
The change of variables  (\ref{a13}) requires the introduction of an additional
term in the action
\begin{equation}
S_m \;=\; {N\over 12}\sum_{\rm sites}{\rm Tr} A^2_i \;+\; O\left( A^4\right)\;.
\label{a15}
\end{equation}
Then following the recipe illustrated in the Gaussian example,
we rewrite the field $A^a(x_s)$ in terms of two new fields $A^a_-(x_-)$
and $A^a_+(x_+)$, whose propagators can easily derived from
those of the Gaussian models, cfr. (\ref{a9}).
We are now ready to perform weak coupling calculations.

Given the free energy per site
\begin{equation}
F(\beta)\;=\; {1\over n_{\rm s} \,N^2 }
\ln \int \prod_x {\rm d}U(x)\exp(-S_{\rm h})\;,
\end{equation}
where $n_s$ is the number of sites,
the internal energy (per link) can be obtained by
\begin{equation}
E\;=\; 1\,-\,{1\over 3}{{\rm d}F(\beta)\over {\rm d}\beta}\;.
\end{equation}
The internal energy up to two loops is given by Eq.~(\ref{energyex}).

In order to evaluate the ratio between the $\Lambda$-parameters
of the ${\overline {MS}}$ renormalization scheme and the honeycomb
lattice regularization, we calculated the correlation function
\begin{equation}
G(T,x_+-y_+)\;=\; {1\over N}
\langle \;{\rm Re} \;{\rm Tr} \,[ U(x_+) U(y_+)^\dagger ]\;\rangle\;\;\;.
\label{a16}
\end{equation}
In the $x$-space we obtained (neglecting $O(a)$ terms)
\begin{equation}
G(T,x,a)\;=\; 1\;+\;{N^2-1\over 2N} \,T \,F(a/x)\;+\;
O\left( T^2 \right)\;,
\label{a17}
\end{equation}
where
\begin{equation}
F(a/x)\;=\; {1\over 2\pi} \left( \ln {a\over x} - \gamma_E - \ln 2\right)\;\;.
\label{a18}
\end{equation}
In the $p$ space
\begin{equation}
\widetilde{G}(T,p,a)\;=\;{N^2-1\over 2N} {T\over p^2}\;\left[
1\;+\; {N^2-2\over 4N}T \left( D(ap)+{1\over 3\sqrt{3}}\right)\;+
O\left( T^2 \right)\right]\;,
\label{a19}
\end{equation}
where
\begin{equation}
D(ap)\;=\; {1\over 2\pi} \left(\ln ap - 2\ln 2\right)\;.
\label{a20}
\end{equation}
The above results required, beside the relation (\ref{a11}),  the calculation
of the following integrals:
\begin{equation}
\int_{-{2\pi\over 3}}^{{2\pi\over 3}}{dk_1\over 2\pi}
\int_{-{\pi\over \sqrt{3}}}^{{\pi\over \sqrt{3}}}{dk_2\over 2\pi}
\;{e^{ikx}-1\over \Delta(k)}\;=\; F(a/x) \;+\; O\left( {a\over x}\right)\;,
\label{a21}
\end{equation}
\begin{equation}
\int_{-{2\pi\over 3}}^{{2\pi\over 3}}{dk_1\over 2\pi}
\int_{-{\pi\over \sqrt{3}}}^{{\pi\over \sqrt{3}}}{dk_2\over 2\pi}
\;{\Delta(p)-\Delta(k)-\Delta(k+p)\over \Delta(k)\Delta(k+p)}\;=\;
2D(ap)\;+\;O\left( ap\right)\;,
\label{a22}
\end{equation}
where the extremes of integration are chosen to cover the appropriate
Brillouin zone, which can be determined from the finite lattice
momenta (\ref{a4}).

The next step consists in determining the
renormalized functions $Z^{\overline {MS}}_t(T,a\mu)$
and $Z^{\overline {MS}}_U(T,a\mu)$ that satisfy the equations
\begin{eqnarray}
&&G_R^{\overline {MS}}(t,x,\mu) \;=\; Z^{\overline
{MS}}_U(T,a\mu)^{-1}\;G(T,x,a)\;\;,\nonumber \\
&&T=Z^{\overline {MS}}_t(T,a\mu)\,t\;\;,
\label{renormG_dim}
\end{eqnarray}
where  $t$ and $G_R^{\overline {MS}}(t,x,\mu)$
are respectively the coupling and
the correlation function renormalized in the ${\overline {MS}}$ scheme.
In the ${\overline {MS}}$ renormalization scheme we have~\cite{chiral2}
\begin{eqnarray}
&&G_R^{\overline {MS}}(t,x\mu=2e^{-\gamma_E})\;=\;1\;+\;O(t^3)\;\;,\nonumber
\\ &&\widetilde{G}_R^{\overline {MS}}(t,{p\over \mu}=1)\;=\;
{N^2-1\over 2N}{t\over p^2}\,\left[ 1\,+\,O\left(t^2\right)\right]\;\;.
\label{renormG}
\end{eqnarray}
Then by imposing Eqs.\ (\ref{renormG_dim}) we obtain
\begin{equation}
Z^{\overline {MS}}_t(T,a\mu)\;=\;
1\;+\;T\,{N\over 8\pi}\,(\ln a\mu \,+\,d)\;+\;O\left(T^2\right)\;,
\label{a25}
\end{equation}
where
\begin{equation}
d\;=\;-2\ln 2\;-\;{N^2-2\over N^2}\,{2\pi\over 3\sqrt{3}}\;.
\label{a26}
\end{equation}
The constant $d$ determines the ratio
$\Lambda_{\overline {MS}}/\Lambda_{\rm h}$, indeed
\begin{equation}
{\Lambda_{\overline {MS}}\over \Lambda_{\rm h}}\;=\; e^{-d}\;=\;
4\exp \left( {N^2-2\over N^2}\,{2\pi\over 3\sqrt{3}} \right)\,\;.
\label{a27}
\end{equation}

For the interested reader we mention that Eqs.~(\ref{a21}-\ref{a22})
may be derived from the following exact result
\begin{eqnarray}
&&\int_{-{2\pi\over 3}}^{{2\pi\over 3}}{dk_1\over 2\pi}
\int_{-{\pi\over \sqrt{3}}}^{{\pi\over \sqrt{3}}}{dk_2\over 2\pi}
\;{1\over \Delta(k)+m^2\left( 1+{m^2\over 8}\right)}\;=\;\nonumber \\
&&{1\over 2\pi} \left( 1+{3\over 8}m^2\right)^{-3/2}
\left( 1+{1\over 8}m^2\right)^{-1/2}
K\left[ \left(1+{1\over 4}m^2\right)^{1/2}
\left(1+{3\over 8}m^2\right)^{-3/2}\left(1+{1\over
8}m^2\right)^{-1/2}\right]\;.
\label{a28}
\end{eqnarray}



\figure{Specific heat vs. $\beta$.
The dashed and solid lines represent the plain strong coupling series and its
resummation.
The estimate of the critical $\beta$ is indicated by vertical dotted lines.
When error bars are not visible, they are smaller than the symbol size.
\label{C_largeN}}

\figure{Magnetic susceptibility vs. $\beta$.
The dashed and solid lines represent the plain strong coupling series and its
resummation
respectively.
The estimate of the critical $\beta$ is indicated by the vertical dotted lines.
\label{magsusc}}

\figure{$\mu_{\rm s}/\mu_{\rm d}$ vs. $\xi_G\equiv 1/M_G$.
\label{rotinv}}

\figure{$\mu_{\rm s}/M_G$ vs. $\xi_G\equiv 1/M_G$.
The dashed lines represents the continuum limit result from Monte Carlo data.
\label{scaling}}

\figure{Asymptotic scaling test by using strong-coupling estimates.
The dotted line represents the exact result (\ref{mass-lambda}).
\label{asyscaling}}

\figure{$\mu_1/\mu_2$ vs. $\xi_G\equiv 1/M_G$ for the honeycomb lattice.
\label{rotinvex}}

\figure{$\mu_1/M_G$ vs. $\xi_G\equiv 1/M_G$ for the honeycomb lattice.
The dashed lines represents the continuum limit result from Monte Carlo data.
\label{scalingex}}

\figure{Asymptotic scaling test for the honeycomb lattice
by using strong-coupling estimates.
The dotted line represents the exact result (\ref{mass-lambdaex}).
The full line corresponding to the ratio $M_G/\Lambda_{{\rm h},E}$
was constructed by resumming the involved strong coupling series.
\label{asyscalingex}}


\begin{table}
\caption{Resummation of the strong coupling series of the specific heat.
We analyze the series of $\beta^{-2}C$ expressed in terms of $\beta^2$,
cf. Eq.~(\ref{Cseries}), for which $M=8$. We report the first singularity in
the real axis,
$\beta_0\equiv \sqrt{x_0}$, and the corresponding exponent versus
$L,K$, and $J$.
}
\label{Cresum}
\begin{tabular}{rrrr@{}lr@{}l}
\multicolumn{1}{r}{$L$}&
\multicolumn{1}{r}{$K$}&
\multicolumn{1}{r}{$J$}&
\multicolumn{2}{c}{$\beta_0$}&
\multicolumn{2}{c}{$\gamma$}\\
\tableline \hline
2 & 2 & 2 & 0&.30598 & 0&.252 \\
2 & 3 & 1 & 0&.30566 & 0&.227 \\
2 & 1 & 3 & 0&.30586 & 0&.245 \\
2 & 0 & 4 & 0&.30563 & 0&.228 \\
2 & 4 & 0 & 0&.30569 & 0&.228 \\
3 & 2 & 1 & 0&.30697 & 0&.280 \\
3 & 1 & 2 & 0&.30591 & 0&.250 \\
3 & 3 & 0 & 0&.30568 & 0&.228 \\
3 & 0 & 3 & 0&.30619 & 0&.277 \\
4 & 1 & 1 & 0&.30508 & 0&.183 \\
4 & 2 & 0 & 0&.30475 & 0&.166 \\
4 & 0 & 2 & 0&.30570 & 0&.233 \\
5 & 1 & 0 & 0&.30562 & 0&.222 \\
5 & 0 & 1 & 0&.30588 & 0&.241 \\
6 & 0 & 0 & 0&.30564 & 0&.225 \\\hline
\end{tabular}
\end{table}

\begin{table}
\caption{Analysis of the series of ${\rm d}\ln\chi/ {\rm d}\beta$ and
${\rm d}\ln \beta M^2_G/ {\rm d}\beta$.
For some set of $L,K$ and $J$ we report the range of values of $\gamma$
corresponding to the range of zero values $0.3055-0.3061$.}
\label{chiresum}
\begin{tabular}{crrrr}
\multicolumn{1}{r}{}&
\multicolumn{1}{r}{$L$}&
\multicolumn{1}{r}{$K$}&
\multicolumn{1}{r}{$J$}&
\multicolumn{1}{c}{$\gamma$}\\
\tableline \hline
${{\rm d}\ln\chi\over {\rm d}\beta}$
                                     &  4 & 5 & 4 & 0.20-0.24 \\
                                     &  4 & 4 & 5 & 0.28-0.31 \\
                                     &  5 & 4 & 4 & 0.23-0.27 \\
                                     &  5 & 5 & 3 & 0.22-0.26 \\
                                     &  6 & 3 & 4 & 0.22-0.26 \\
                                     &  6 & 4 & 3 & 0.23-0.27 \\\hline

${{\rm d}\ln (\beta M^2_G)\over {\rm d}\beta}$
                           &  4 & 4 & 4 & 0.23-0.26 \\
                           &  5 & 4 & 3 & 0.26-0.30 \\
                           &  5 & 3 & 4 & 0.07-0.12 \\
                           &  6 & 3 & 3 & 0.10-0.15 \\
\end{tabular}
\end{table}

\begin{table}
\caption{Resummation of the $24^{\rm th}$ order strong coupling series of the
specific heat
for the honeycomb lattice.
We analyze the series of $\beta^{-2}C$ expressed in terms of $\beta^2$,
cf. Eq.~(\ref{Cseriesex}), for which $M=12$.
We report the first singularity in the real axis,
$\beta_0\equiv \sqrt{x_0}$, and the corresponding exponent versus
$L,K$, and $J$.
}
\label{Cresumex}
\begin{tabular}{rrrr@{}lr@{}l}
\multicolumn{1}{r}{$L$}&
\multicolumn{1}{r}{$K$}&
\multicolumn{1}{r}{$J$}&
\multicolumn{2}{c}{$\beta_0$}&
\multicolumn{2}{c}{$\gamma$}\\
\tableline \hline
3 & 4 & 3 & 0&.43386 & 0&.162 \\
3 & 3 & 4 & 0&.43389 & 0&.165 \\
4 & 3 & 3 & 0&.43393 & 0&.167 \\
4 & 4 & 2 & 0&.43398 & 0&.171 \\
4 & 2 & 4 & 0&.43381 & 0&.161 \\
5 & 3 & 2 & 0&.43387 & 0&.163 \\
5 & 2 & 3 & 0&.43415 & 0&.185 \\
5 & 4 & 1 & 0&.43397 & 0&.171 \\
5 & 1 & 4 & 0&.43495 & 0&.240 \\
6 & 2 & 2 & 0&.43312 & 0&.101 \\
\end{tabular}
\end{table}

\begin{table}
\caption{Analysis of the series of ${\rm d}\ln\chi/ {\rm d}\beta$ ($19^{\rm
th}$ order)
and ${\rm d}\ln \beta M^2_G/ {\rm d}\beta$ ($18^{\rm th}$ order) for the
honeycomb lattice.
For some set of $L,K$ and $J$ we report the range of values of $\gamma$
corresponding to the range of zero values $0.4338-0.4340$.}
\label{chiresumex}
\begin{tabular}{crrrr}
\multicolumn{1}{r}{}&
\multicolumn{1}{r}{$L$}&
\multicolumn{1}{r}{$K$}&
\multicolumn{1}{r}{$J$}&
\multicolumn{1}{c}{$\gamma$}\\
\tableline \hline
${{\rm d}\ln\chi\over {\rm d}\beta}$
                                     &  6 & 6 & 6 & 0.15-0.16 \\
                                     &  6 & 7 & 5 & 0.14-0.15 \\
                                     &  6 & 5 & 7 & 0.14-0.15 \\
                                     &  7 & 5 & 6 & 0.15-0.16 \\
                                     &  7 & 6 & 5 & 0.30-0.30 \\
                                     &  8 & 5 & 5 & 0.15-0.16 \\\hline

${{\rm d}\ln (\beta M^2_G)\over {\rm d}\beta}$
                           &  6 & 6 & 5 & 0.21-0.22 \\
                           &  6 & 5 & 6 & 0.22-0.23 \\
                           &  7 & 5 & 5 & 0.21-0.22 \\
                           &  8 & 4 & 5 & 0.17-0.18 \\
                           &  8 & 5 & 4 & 0.22-0.24 \\
\end{tabular}
\end{table}

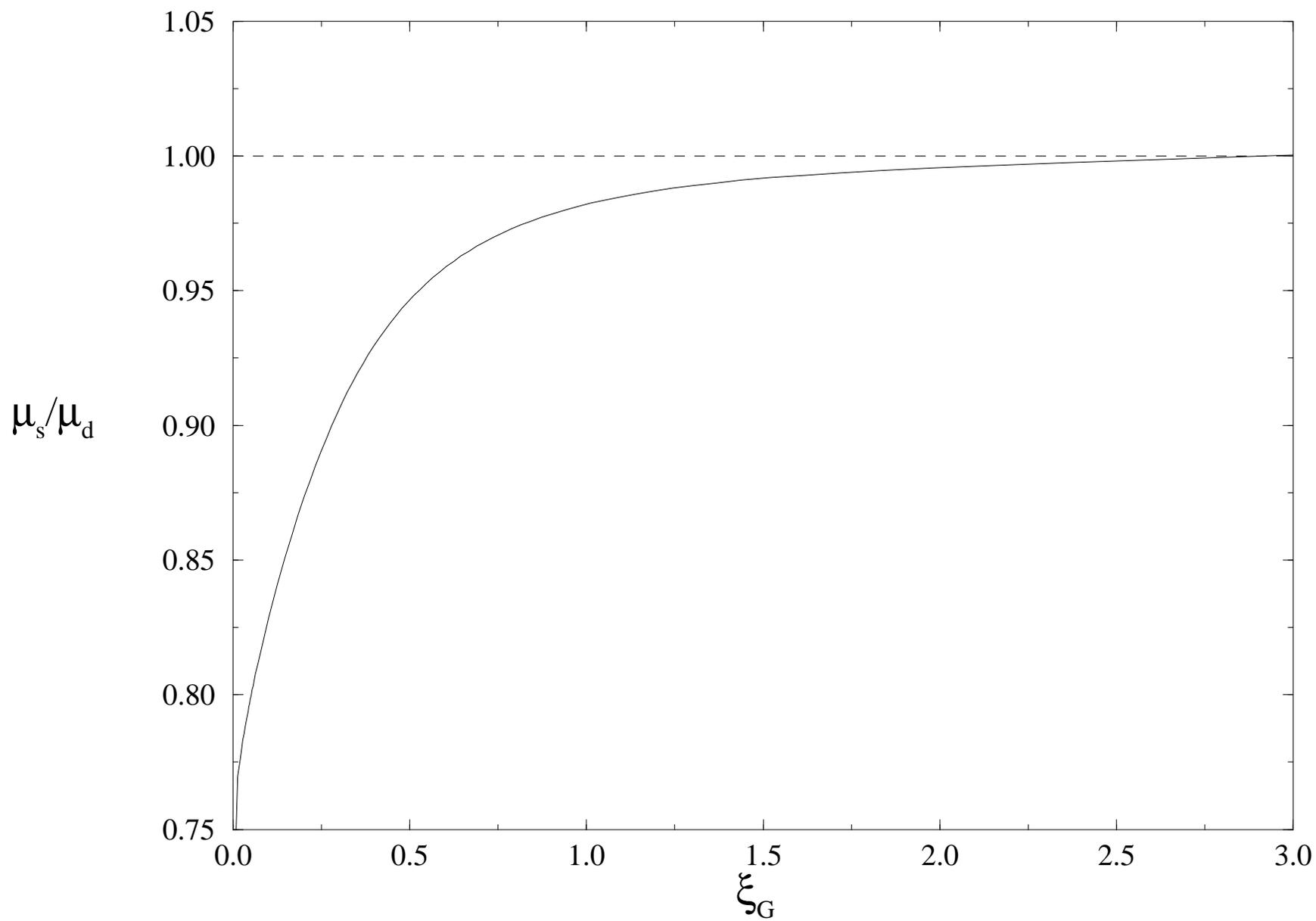
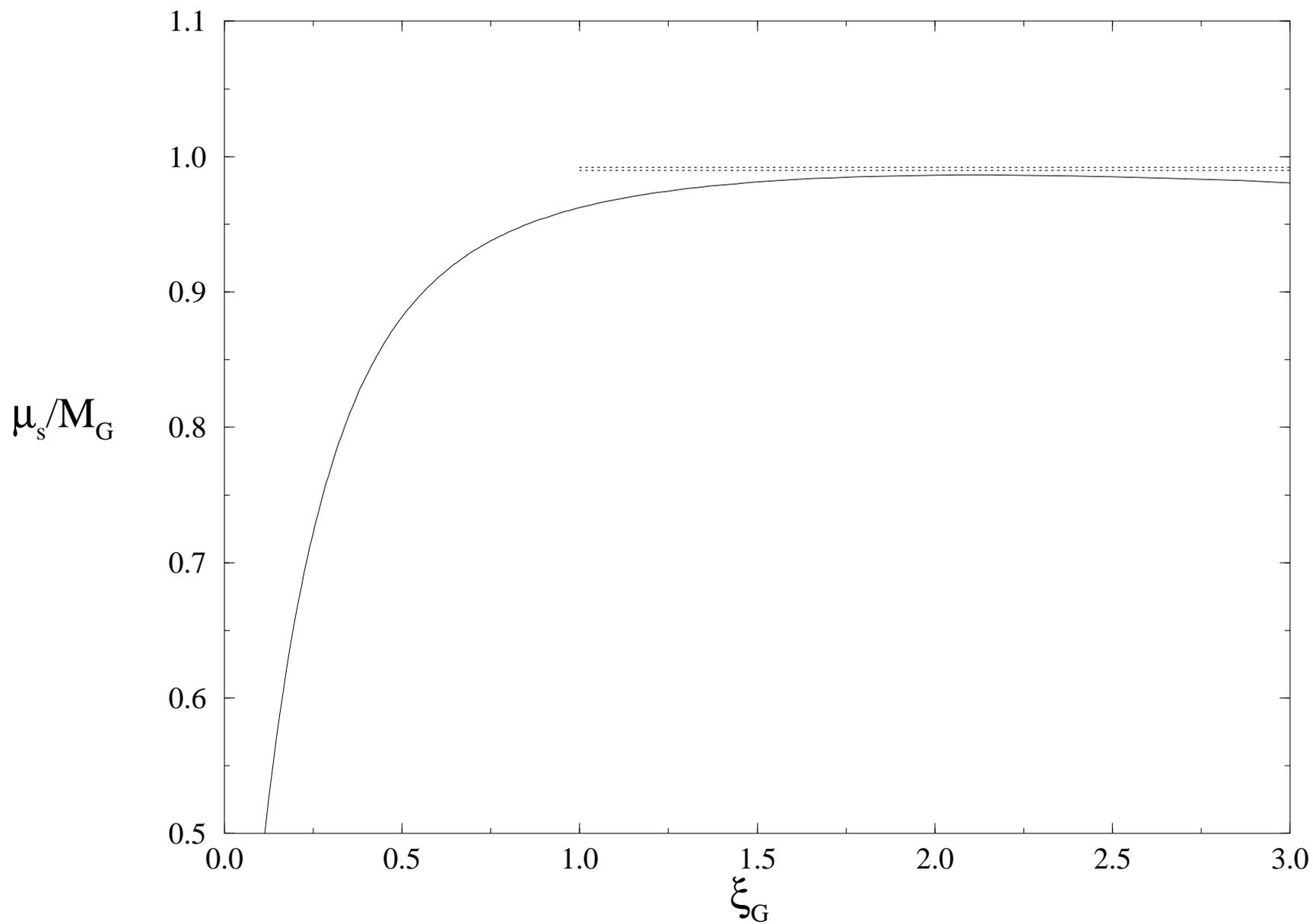
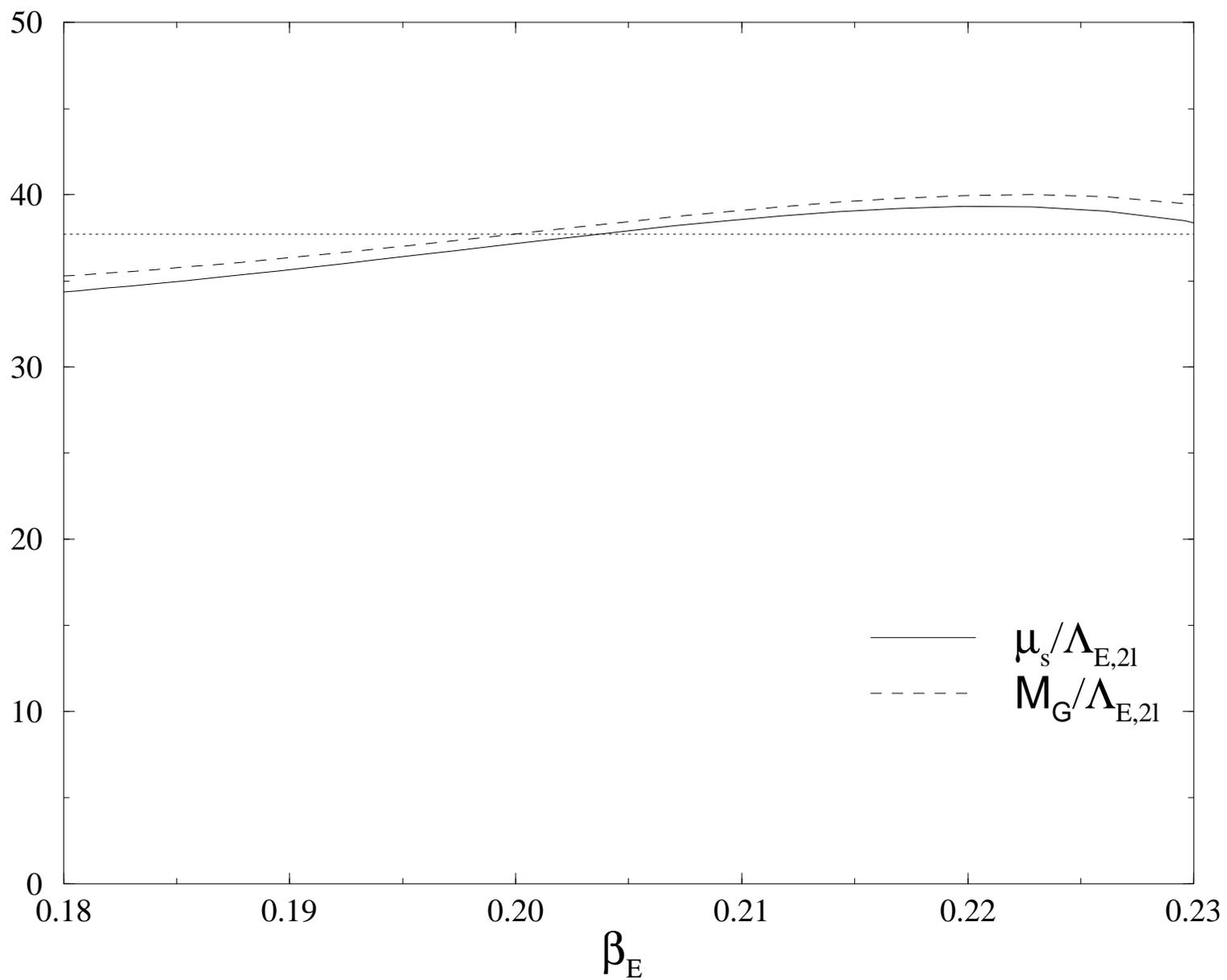
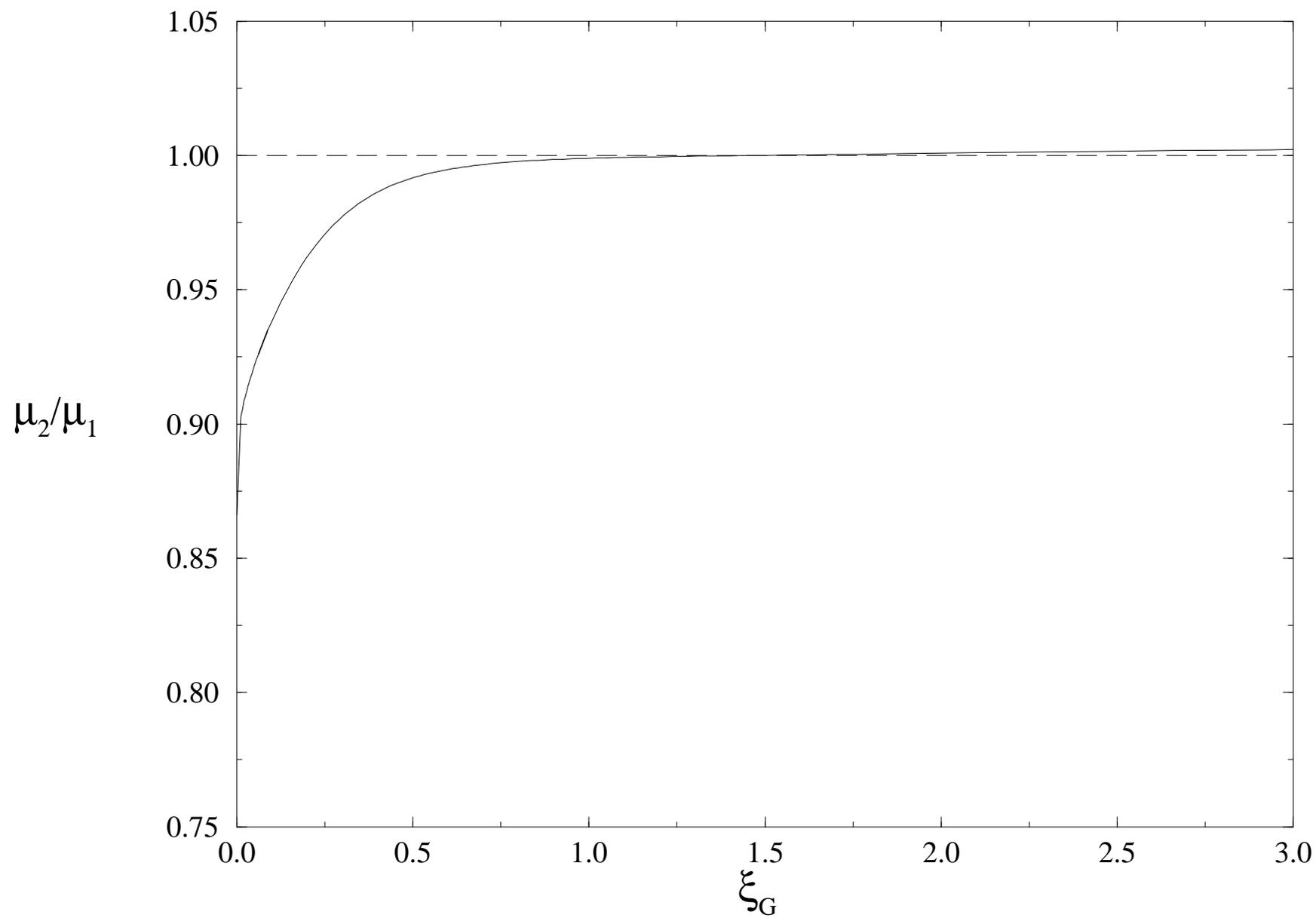
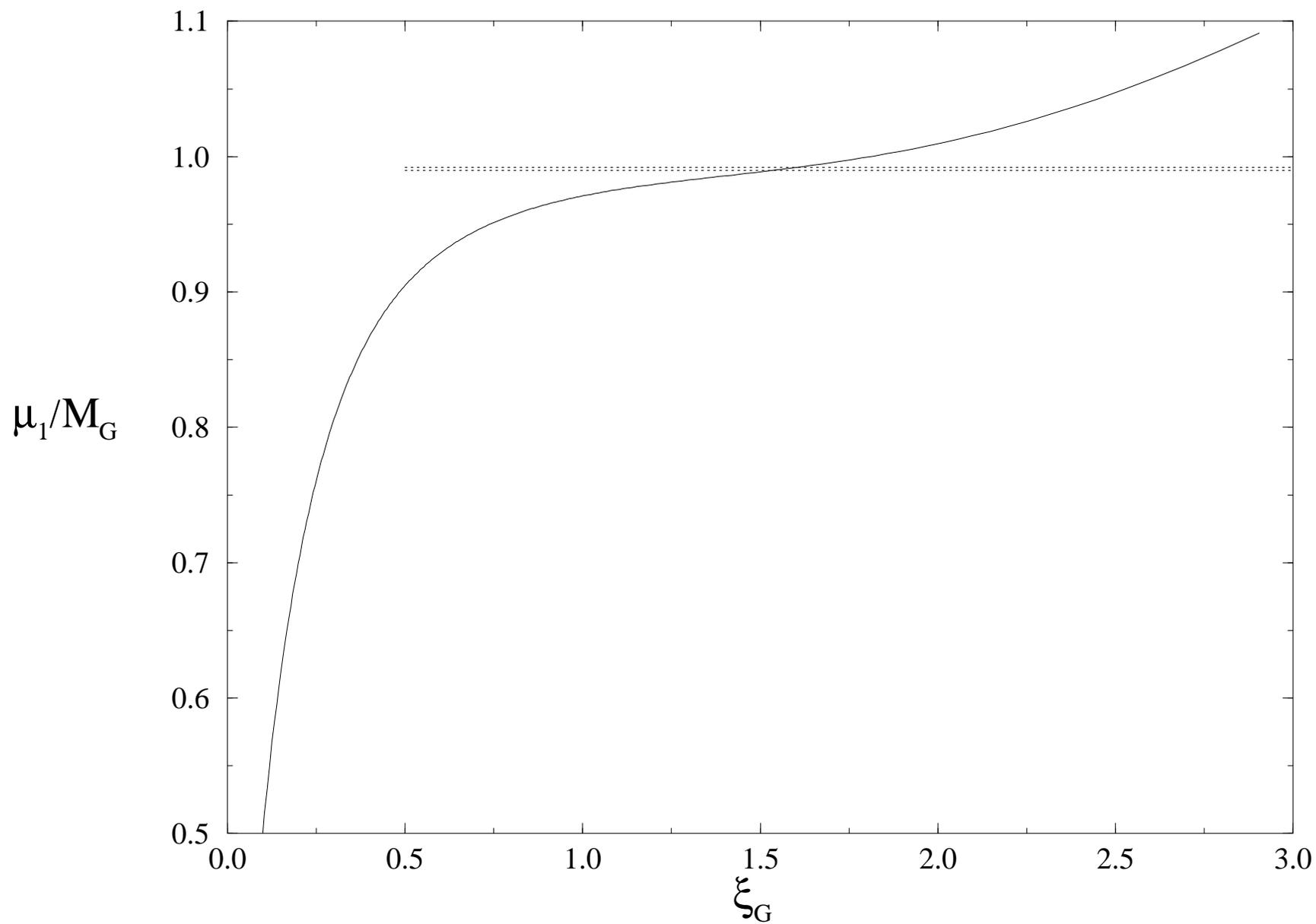
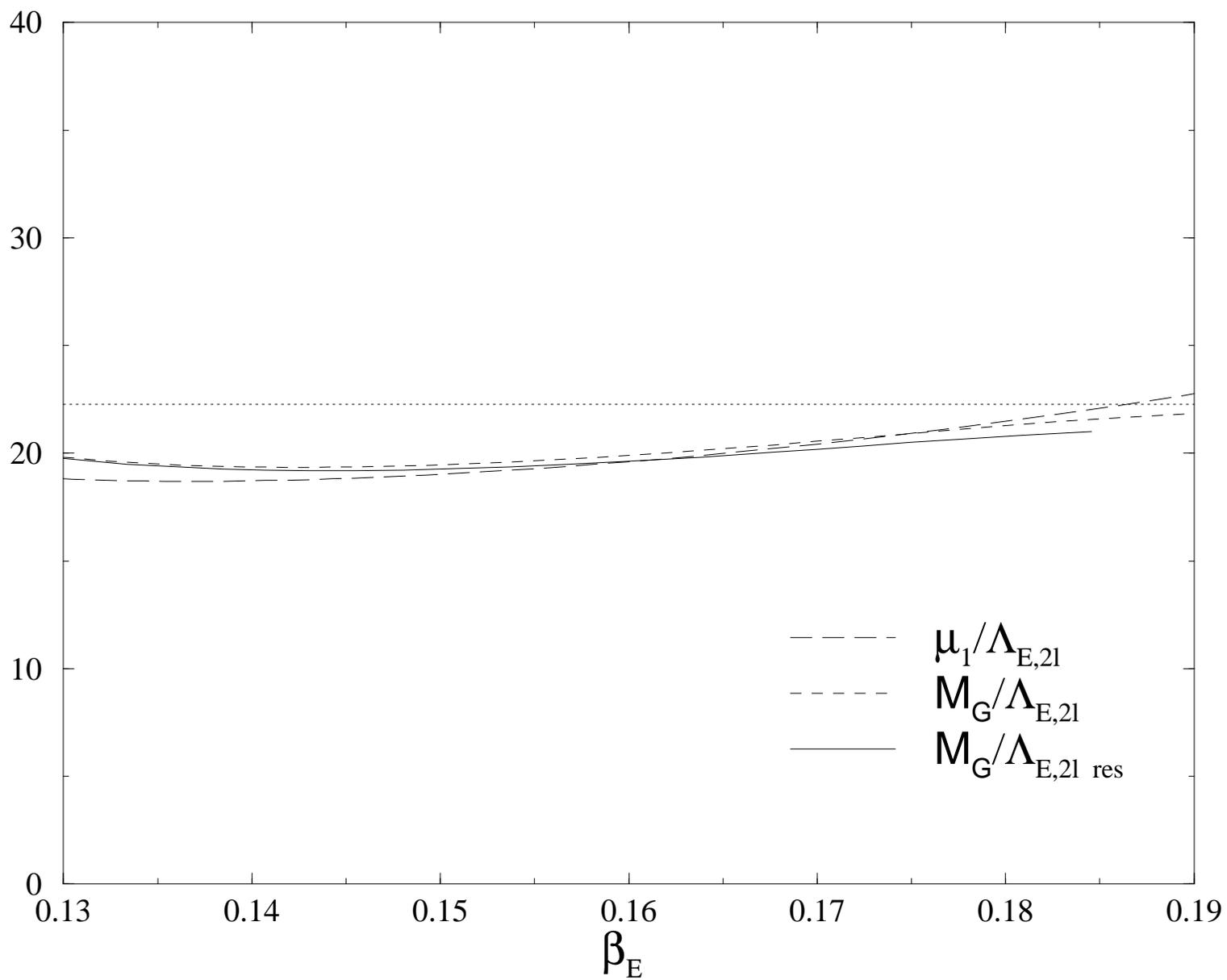
\end{document}